\documentclass[preprint,amsmath,showpacs]{revtex4}
\usepackage{graphicx}
\usepackage{dcolumn}
\usepackage{bm}
\usepackage{epstopdf}

\def\bee{\begin{eqnarray}}
\def\eee{\end{eqnarray}}

\begin{document}

\title{Determination of the Source Flavor Ratio of Ultrahigh Energy Neutrinos }

\author{T.C. Liu$^{1,3}$}

\email{tcliu@ntu.edu.tw}

\author{M.A. Huang$^{2,3}$}

\email{mahuang@nuu.edu.tw}

\author{Guey-Lin Lin$^{1,3}$}

\email{glin@mail.nctu.edu.tw}

\affiliation{$^{1}$Institute of Physics, National Chiao-Tung
University, Hsinchu 30010, Taiwan}

\affiliation{$^{2}$Department of Energy and Resources, National
United University, Maio-Li 36003, Taiwan}

\affiliation{$^{3}$Leung Center for Cosmology and Particle
Astrophysics, National Taiwan University, Taipei 106, Taiwan}

\begin{abstract}
We discuss the reconstruction of neutrino flavor neutrino at a
distant source in the very high energy regime. This reconstruction
procedure is relevant to the confirmation of detecting cosmogenic
neutrinos, for example. To facilitate such a reconstruction, it is
imperative to achieve effective flavor discriminations in
terrestrial neutrino telescopes. We note that, for energies beyond
few tens of PeV, a tau-lepton behaves like a track similar to a
muon. Hence, while it is rather challenging to separate $\nu_{\mu}$
from $\nu_{\tau}$ in this case, one can expect to isolate $\nu_e$
from the rest by a distinctive shower signature. We present the
result of flavor ratio reconstruction given the anticipated
accuracies of flavor measurement in neutrino telescopes and current
uncertainties of neutrino mixing parameters. It is shown that the
further separation between $\nu_{\mu}$ and $\nu_{\tau}$ events does
not improve the flavor reconstruction due to the approximate
$\nu_{\mu}-\nu_{\tau}$ symmetry.

\end{abstract}

\maketitle

\section{Introduction}
High energy neutrinos with different flavor ratios can be produced
from various astrophysical environments or production processes. For
instance, a typical hadronic interaction produces neutrinos through
$\pi^{\pm}\rightarrow\mu^{\pm}\rightarrow e^{\pm}$ decay chains and
generates a neutrino flavor ratio,
$\phi_{0}(\nu_{e}):\phi_{0}(\nu_{\mu}):\phi_{0}(\nu_{\tau})=1:2:0$
\cite{Pakvasa1}. This type of source shall be referred to as pion
source hereafter. When the source is hidden behind a strong field
\cite{Kachelriess} or a dense matter \cite{Rachen}, the muon
produced by the pion decay may lose a significant amount of its
energy before its decay \cite{muonenergyloss}. This type of source
is referred to as muon-damped source since the neutrinos produced by
the decay of secondary muon give negligible contribution to the
total neutrino flux. Hence the resulting neutrino flavor ratio is
$\phi_{0}(\nu_{e}):\phi_{0}(\nu_{\mu}):\phi_{0}(\nu_{\tau})=0:1:0$.

The flavor ratio of neutrinos are altered by oscillation effects
while neutrinos propagate from the source to the detector on the
Earth. Since the distance from an astrophysical source to the Earth
is generally much longer than the neutrino oscillation length, the
observed neutrino flavor ratio is fully determined by three neutrino
mixing angles and the CP phase. In this work, we aim to reconstruct
the neutrino flavor ratio at the source from the observed flavor
ratio on the Earth. As we shall see later, the method for such a
reconstruction depends on the neutrino energies. We shall focus our
discussion on the highest energy regime, which is an extension to
our earlier work \cite{Lai}.

\section{Classification of neutrino events at different energies}
\begin{figure}
\includegraphics[scale=0.4]{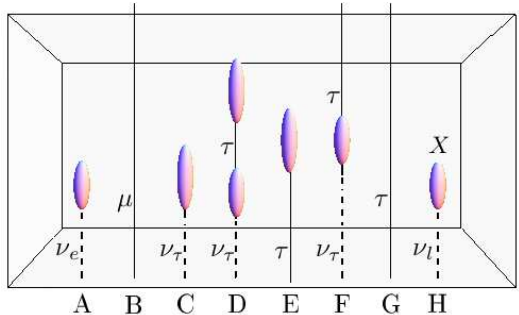}

\caption{Different types of neutrino-induced events. Dashed lines and solid
lines correspond to paths of neutrinos and leptons respectively. The
ellipsoids are showers. The energy range suitable for detecting each type
of event is listed in Table~\ref{tab:behave}.}

\label{fig:TrackandShower}
\end{figure}

 Neutrinos interact
with matters to produce observable signals. The major channel for
such interactions is the charged-current (CC) interaction,
$\nu_{l}+N\longrightarrow l+X$, where $l$ is the lepton associated
with $\nu_{l}$ and $X$ denotes the hadronic states. The sub-dominant
channel is the neutral-current interaction (NC),
$\nu_{l}+N\longrightarrow\nu_{l}+X$. In
Fig.~\ref{fig:TrackandShower} and Table~\ref{tab:behave}, we
summarize different types of neutrino induced events and their
detectable energy ranges.

\begin{table*}
\begin{tabular}{>{\raggedright}p{2in}lll}
\hline particle  & major processes  & signal type  & symbol in
Fig.\ref{fig:TrackandShower}\tabularnewline \hline e  & EM shower  &
shower  & A\tabularnewline $\mu$  & energy loss  & track  &
B\tabularnewline $\tau$($E_{\nu}<3.3$ PeV)  & CC int. and
$\tau$-decay  & shower  & C\tabularnewline $\tau$($3.3$
PeV$<E_{\nu}<33$ PeV)  & CC int. and $\tau$-decay  & 2 separate
showers  & D (double-bang event)\tabularnewline $\tau$($E_{\nu}>3.3$
PeV)  & energy loss and decay  & track and shower  & E (lollipop
event)\tabularnewline $\tau$($E_{\nu}>3.3$ PeV)  & CC int. and
energy loss  & shower and track  & F (inverted lollipop
event)\tabularnewline $\tau$($E_{\nu}>33$ PeV)  & energy loss  &
track  & G\tabularnewline $X$  & hadron shower  & shower  &
H\tabularnewline \hline
\end{tabular}

\caption{Different types of neutrino induced events.}

\label{tab:behave}
\end{table*}

Type-A event in Fig.~\ref{fig:TrackandShower} is an electron
production through $\nu_{e}$ CC interaction. The electron has a
large interaction cross section with the medium and produces a
shower within a short distance from its production point.
Type-B event is a muon
produced by $\nu_{\mu}$ CC interaction. Contrary to the electron, a
muon can travel a long distance in the medium before it loses all
its energy or decays. The muon range in ice is more than $10$ km for
$E_{\nu}=1$ PeV ($10^{15}$ eV). Hence, above this energy, there is
hardly any decay of muon occurring within the fiducial volume of the
detector, which is about a few km$^{3}$. A muon does, however, lose
a small fraction of its energy and emits dim lights so that only
those optical detectors which are near to the muon track can be
triggered. As a result, a muon produces a track-like signal.

The $\nu_{\tau}$-induced events are listed as types C-G where the
tau lepton produced by $\nu_{\tau}$ CC interaction behaves differently
at different energies for a fixed detector design. For a neutrino
telescope such as IceCube \cite{Icecube}, the distance between each
string of optical detectors is 125 m, which corresponds to the decay
length of a $2.5$ PeV tau lepton. Such a tau lepton could be produced
by the CC interaction of a $\nu_{\tau}$ with $E_{\nu}=3.3$ PeV.
Therefore, for a $\nu_{\tau}$ with an energy significantly smaller
than this, the separation between the first hadronic shower produced
by CC interaction and the second shower produced by the tau-lepton
decay is too small to be resolved. Such an event is classified as
type C. For $E_{\nu}>3.3$ PeV, one can resolve the above double-bang
event (classified as type D) until the separation of two showers exceeds
the effective size of the detector. Such a size is estimated to be
the sum of IceCube dimension ($\approx1$ km) and two extinction lengths
of optical photons in ice ($\approx250$ m), which corresponds to
the decay length of a tau lepton with $E_{\tau}=25$ PeV. The average
energy of $\nu_{\tau}$ capable of producing such a tau lepton is
around $33$ PeV.
Hence the configuration of IceCube detector determines
the observable energy range for the
double bang event to be $3.3\ {\rm PeV}<E_{\nu}<33\ {\rm PeV}$
\cite{Learned95,Athar,Cowen}. For an under-sea experiment, such
as KM3Net \cite{KM3net}, the observable energy range for the double
bang event is similar.

Type-E event is referred to as the lollipop event. In such an event,
a high energy tau lepton enters the detector and decays within it,
producing a track signal followed by a shower. The probability for
observing a lollipop event increases with the neutrino energy, and
it is about $5\times10^{-4}$ for $E_{\nu}=1$ EeV \cite{Beacom}.
Type-F event is the inverted lollipop which consists of a hadronic
shower from $\nu_{\tau}$ CC interaction and a subsequent tau-lepton
track. Both muons and tau leptons produce inverted lollipop events
and it is not easy to separate them. Type-G event is a through-going
tau-lepton track which is produced by $\nu_{\tau}$ CC interaction
with $E_{\nu}>33$ PeV. Finally type-H event is induced by the
neutral-current interaction.

\section{Flavor Discriminations for different neutrino energies}
Although our focused energy range is at $E_{\nu}> 33$ PeV, it is
helpful to review the flavor discrimination in the lower energy
ranges.
\subsection{Flavor discrimination for $E_{\nu}<3.3$ PeV}
In this energy range, type-C events can not be separated from type-A
and type-H events since the two showers in type-C events can not be
resolved. Hence one can only distinguish muon track event (type B)
from shower events (type-A, C and H). In IceCube, such a distinction
can be done effectively \cite{Beacom}, which is useful for deducing
the flux ratio $R^{{\rm
I}}=\phi(\nu_{\mu})/(\phi(\nu_{e})+\phi(\nu_{\tau}))$
\cite{KachelriesswithR}. We note that $\phi(\nu_{\mu})$ in the
numerator contributes to both type-B and type-H events. On the other
hand, $\phi(\nu_{e})$ and $\phi(\nu_{\tau})$ contribute equally to
type-H events if $\phi(\nu_{e})=\phi(\nu_{\tau})$. Furthermore, type-A and type-C events occur with the
same scattering cross section \cite{comment}. This explains the flux combination
$\phi(\nu_{e})+\phi(\nu_{\tau})$ appearing in the denominator.

\subsection{Flavor discrimination for $3.3\ {\rm PeV}<E_{\nu}<33\ {\rm PeV}$}
In this energy range, one can detect the type-D and type-E
$\nu_{\tau}$ events (double bang and lollipop). Hence it is also
possible to deduce the flux ratio $S^{{\rm
I}}\equiv\phi(\nu_{e})/\phi(\nu_{\tau})$ \cite{WinterithR} in
addition to $R^{\rm I}$. However, the double bang and lollipop
events are both rare so that the error associated with $S^{\rm I}$
is large. In an earlier paper \cite{Lai}, we demonstrated that a
large number of events is necessary for lowering down the errors of
$R^{{\rm I}}$ and $S^{{\rm I}}$ to the point that one can
distinguish the pion source from the muon-damped source.

\begin{table}

\begin{tabular}{cc}
\hline Condition I : ${E}_{\nu}<33$ PeV & Condition II :
${E}_{\nu}>33$ PeV\tabularnewline \hline $R^{{\rm
I}}=\phi(\nu_{\mu})/(\phi(\nu_{e})+\phi(\nu_{\tau}))$  & $R^{{\rm
II}}=\phi(\nu_{e})/(\phi(\nu_{\mu})+\phi(\nu_{\tau}))$\tabularnewline
$S^{{\rm I}}=\phi(\nu_{e})/\phi(\nu_{\tau})$  & $S^{{\rm
II}}=\phi(\nu_{\mu})/\phi(\nu_{\tau})$\tabularnewline \hline
\end{tabular}

\caption{Flavor discrimination variables for different energy ranges
}

\label{tab:Condition-set}

\end{table}

\subsection{Flavor discrimination for
$E_{\nu}>33$ PeV} In this high energy regime, the tau-lepton range
becomes long enough so that a tau lepton could pass through the
detector fiducial volume without decaying. In this case, the
tau-lepton loses its energy just like a muon does and the signal
appears like a track event \cite{Bugaev}. Thus, from an experimental
point of view, one should classify such a signature as a track event
(type G). In this energy range, there are also type-E and type-F
events where tau leptons also behave like tracks. It is clear that
the discrimination between tau-lepton and muon tracks are
challenging, i.e, it is non-trivial to measure the parameter
$S^{{\rm II}}=\phi(\nu_{\mu})/\phi(\nu_{\tau})$. However, one can
expect to measure the $\nu_e$ fraction  $R^{{\rm
II}}=\phi(\nu_{e})/(\phi(\nu_{\mu})+\phi(\nu_{\tau}))$ by
identifying the shower signature from the electron. In fact, $\nu_e$
can also be separated from other flavors in neutrino telescopes
based upon radio detection technique \cite{Gerhardt:2010js,Allison:2011wk} due to
Landau-Pomeranchuk-Midal effect
\cite{Landau,Midal,AlvarezMuniz:1999xx}.

As a short summary, we present in Table~\ref{tab:Condition-set} the
appropriate flavor discrimination variables for $E_{\nu}<33$ PeV and
$E_{\nu}>33$ PeV respectively.

\section{The reconstruction of source flavor ratio for $E_{\nu}>33$ PeV}

The evolution of neutrino flavor ratio from the source to the Earth
is given by the probability matrix $P_{\alpha\beta}$ such that
\begin{equation}
\phi(\nu_{\alpha})=\sum P_{\alpha\beta}\phi_{0}(\nu_{\beta}).\label{eq:Pij}
\end{equation}
Each matrix element $P_{\alpha\beta}$ is a function of neutrino
mixing angles and CP violation phase, $\phi(\nu_{\alpha})$ is the
flux of neutrinos of flavor $\alpha$ on the Earth, and
$\phi_{0}(\nu_{\beta})$ is the flux of neutrinos of flavor $\beta$
at the source. We assume the neutrino propagation distance is
sufficiently large so that each $P_{\alpha\beta}$ depends neither on
the neutrino mass-squared differences $\Delta m_{ij}^{2}$ nor on the
neutrino energy $E_{\nu}$.

\subsection{The flavor reconstruction with only $R^{\rm II}$ measured}
If one only measures $R^{\rm II}$, the reconstructed range for
neutrino flavor ratio at the source is given by the following
fitting formula \cite{Blum}
\begin{equation}
\chi^{2}=\left(\frac{R_{{\rm th}}^{{\rm II}}-R_{{\rm exp}}^{{\rm
II}}} {\sigma_{R_{{\rm exp}}^{{\rm II}}}}\right)^{2}+
\underset{_{ij=12,13,23}}{\sum}\chi_{\theta_{ij}}^{2},
\label{chisquare}
\end{equation}
 where
$\sigma_{R_{{\rm exp}}^{{\rm II}}}=(\Delta R^{{\rm II}}/R^{{\rm
II}})R_{{\rm exp}}^{{\rm II}}$. The functions
$\chi_{\theta_{12}}^{2}$ and $\chi_{\theta_{23}}^{2}$ are taken from
Ref.~\cite{Schwetz:2008er} while $\chi_{\theta_{13}}^{2}$ is taken
from Ref.~\cite{Schwetz:2011qt}.  In the above references, the
functional dependence of each $\chi_{\theta_{ij}}^{2}$ on
$\sin^{2}\theta_{ij}$ is given. The suffix {}``th'' indicates the
theoretical predicted values which depend on the source neutrino
flavor ratio and the neutrino mixing angles. The suffix {}``exp''
indicates the experimentally measured values which are generated by
the true neutrino flavor ratio at the source and the best-fit values
of neutrino mixing angles. The best-fit values and the
allowed $1\sigma$ and $3\sigma$ ranges of mixing angles are given by
\begin{equation}
 \sin^{2}\theta_{12}=0.304_{-0.016,0.054}^{+0.022,0.066},
\quad\sin^{2}\theta_{23}=0.5_{-0.06,0.14}^{+0.07,0.17},
\quad\sin^{2}\theta_{13}=0.01_{-0.006}^{+0.009}, \label{angles}
\end{equation}
where the $1\sigma$ range for $\theta_{13}$ used in our analysis is
that associated with normal hierarchy and the $3\sigma$ range for
$\theta_{13}$ in the same mass hierarchy is $\sin^2\theta_{13}\leq
0.035$ \cite{update}. We take the CP phase $\delta=\pi$ for
generating $R_{{\rm exp}}^{{\rm II}}$. We have found that other $\delta$ values do not
produce noticeably different results.
We consider all possible neutrino flavor ratios at the source for
calculating $\chi^2$. Since we have taken $R_{{\rm exp}}^{{\rm II}}$
 as those generated by input true values of initial neutrino flavor
ratios and neutrino mixing parameters, we have $\chi^2_{{\rm
min}}=0$ occurring at these input true values of parameters. The
boundaries for $1\sigma$ and $3\sigma$ ranges of initial neutrino
flavor ratios are given by $\Delta\chi^{2}=2.3$ and
$\Delta\chi^{2}=11.8$ respectively, where
$\Delta\chi^{2}\equiv\chi^{2}-\left(\chi^{2}\right)_{{\rm
min}}=\chi^{2}$ in this analysis.

\begin{figure}
\includegraphics[scale=0.50]{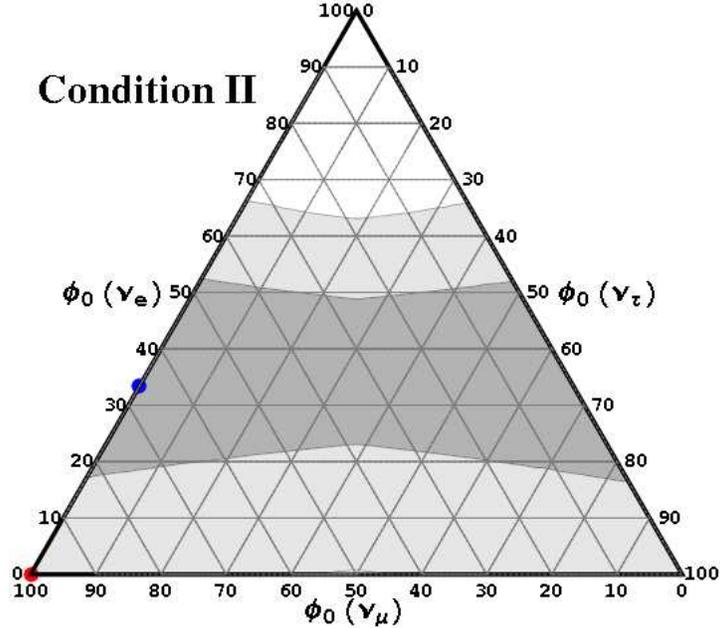}

\caption{Reconstructed flavor ratios for an input pion source with
$\Delta R^{{\rm II}}/R^{{\rm II}}=15\%$. The dark and light shaded
areas denote 1$\sigma$ and 3$\sigma$ ranges respectively.}
\label{fig:PionOnlyR}
\end{figure}

Let us first take the pion source as the input true source and
consider its reconstruction. The accuracy for measuring $R^{{\rm
II}}$ is taken to be $\Delta R^{{\rm II}}/R^{{\rm II}}=15\%$. The
possibility for measuring $R^{\rm II}$ in such an accuracy will be
discussed in the next section. The result for the reconstruction of
neutrino flavor ratio is shown in Fig.~\ref{fig:PionOnlyR}. It is
seen that the muon-damped source can be ruled out at the $1\sigma$
but not on the $3\sigma$ level.
This result shows how well one can discriminate between
cosmogenic neutrino flux (a pion source)
\cite{cosmogenic_early,cosmogenic_update} arising from GZK
interactions \cite{Greisen,ZK} and the high energy tail of
astrophysical neutrino flux (a muon-damped source
\cite{Kashti:2005qa,Lipari:2007su}) when the former type of neutrino
flux is detected.

\begin{figure}[h]
\includegraphics[scale=0.50]{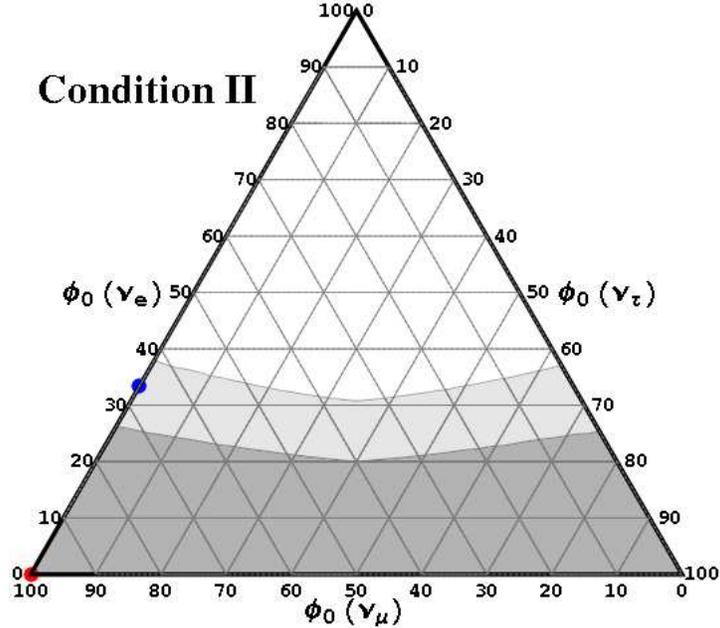}

\caption{Reconstructed flavor ratios for an input muon-damped source
with $\Delta R^{{\rm II}}/R^{{\rm II}}=15\%$. The dark and light
shaded areas denote 1$\sigma$ and 3$\sigma$ ranges respectively.  }
\label{fig:MuonOnlyR}
\end{figure}

We next take the muon-damped source as the input true source. Such a
neutrino flavor ratio occurs at the high energy tail of
astrophysical neutrino flux as just mentioned. Once more we take
$\Delta R^{{\rm II}}/R^{{\rm II}}=15\%$. It is seen from
Fig.~\ref{fig:MuonOnlyR} that the pion source can be ruled out at
the $1\sigma$ but not at the $3\sigma$ level from the flavor reconstruction.

\subsection{The flavor reconstruction with both $R^{\rm II}$ and $S^{\rm II}$ measured}
Although it is very challenging to measure $S^{\rm II}$, it is of
interests to see whether or not such an effort is useful for
improving the reconstruction of neutrino flavor ratio at the source.
Let us take a very optimistic scenario that the error of $S^{\rm
II}$ is comparable to that of $R^{\rm II}$. The statistical analysis
can be performed by adding $S^{\rm II}$ contribution to
Eq.~(\ref{chisquare}).

\begin{figure}
\includegraphics[scale=0.50]{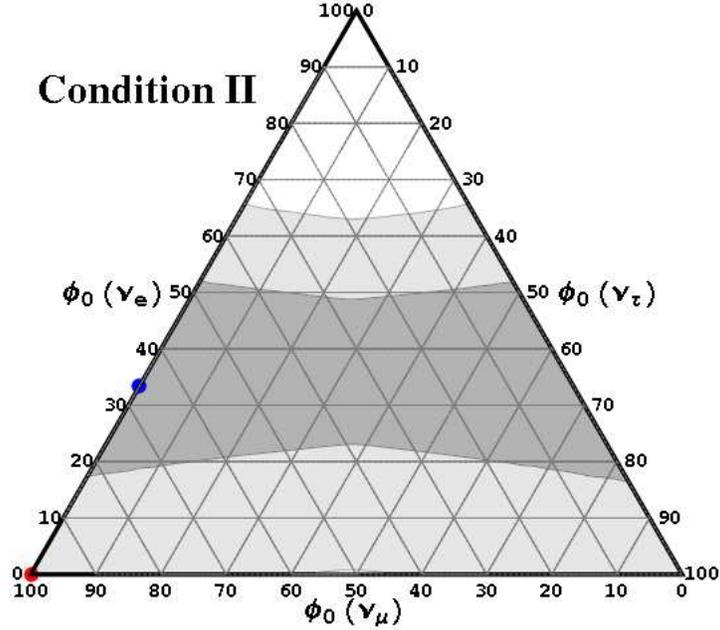}

\caption{Reconstructed flavor ratios for an input pion source with
$\Delta R^{{\rm II}}/R^{{\rm II}}=15\%$ and $\Delta S^{{\rm
II}}/S^{{\rm II}}$ related to the former by Poisson statistics. The
dark and light shaded areas denote the 1$\sigma$ and 3$\sigma$
ranges respectively. The result shown in this figure is comparable
to that in Fig.~\ref{fig:PionOnlyR} where only $R^{\rm II}$ is
measured.}

\label{fig:Pion}
\end{figure}

\begin{figure}
\includegraphics[scale=0.50]{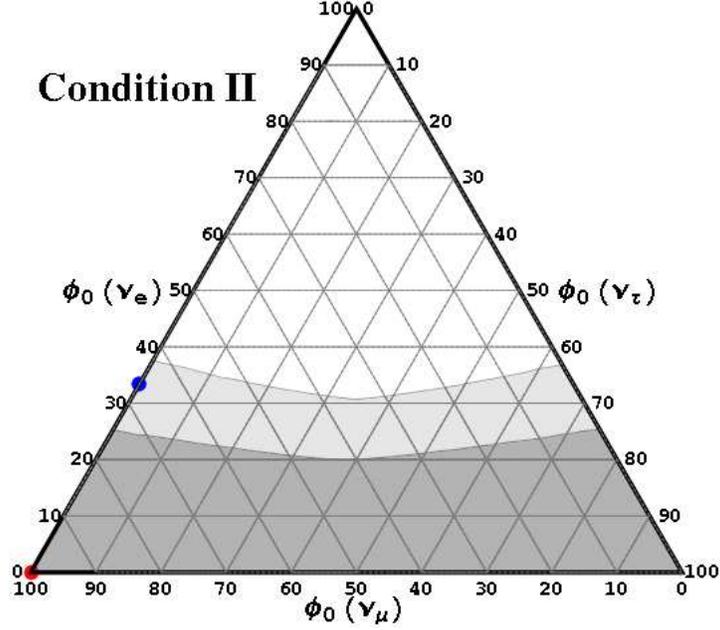}

\caption{Reconstructed flavor ratios for an input muon-damped source
with $\Delta R^{{\rm II}}/R^{{\rm II}}=15\%$ and $\Delta S^{{\rm
II}}/S^{{\rm II}}$ related to the former by Poisson statistics. The
dark and light shaded areas denote 1$\sigma$ and 3$\sigma$ ranges
respectively. The result shown in this figure is comparable to that
in Fig.~\ref{fig:MuonOnlyR} where only $R^{\rm II}$ is measured.}
\label{fig:Muon}
\end{figure}

The reconstructed neutrino flavor ratio for an input pion source is
presented in Fig.~\ref{fig:Pion} where $\Delta R^{{\rm II}}/R^{{\rm
II}}=15\%$ and $\Delta S^{{\rm II}}/S^{{\rm II}}$ related to the
former by Poisson statistics.
This result is in fact comparable to that in
Fig.~\ref{fig:PionOnlyR} where only $R^{\rm II}$ is measured, i.e.,
the further measurement of $S^{\rm II}$ does not tighten the
constraint on the initial neutrino flavor ratio. This can be
understood by the approximate $\nu_{\mu}-\nu_{\tau}$ symmetry
\cite{Balantekin:1999dx,Harrison:2002et} which makes $S^{\rm II}$
always rather close to unity no matter what the initial neutrino
flavor ratio is. Hence the measurement of $S^{\rm II}$ does not help
to constrain the initial neutrino flavor ratio. We can demonstrate
this further by taking the muon-damped source as the input true
source. With $\Delta R^{{\rm II}}/R^{{\rm II}}=15\%$ and $\Delta
S^{{\rm II}}/S^{{\rm II}}$ related to the former by Poisson
statistics, the result for the flavor reconstruction is presented in
Fig.~\ref{fig:Muon}.
Comparing with Fig.~\ref{fig:MuonOnlyR}, one can also see that
the measurement of $S^{\rm II}$ does not help to constrain the
initial neutrino flavor ratio.
\subsection{Comparison to the flavor reconstruction for $E_{\nu}< 33$ PeV }
\begin{figure}
\includegraphics[scale=0.35]{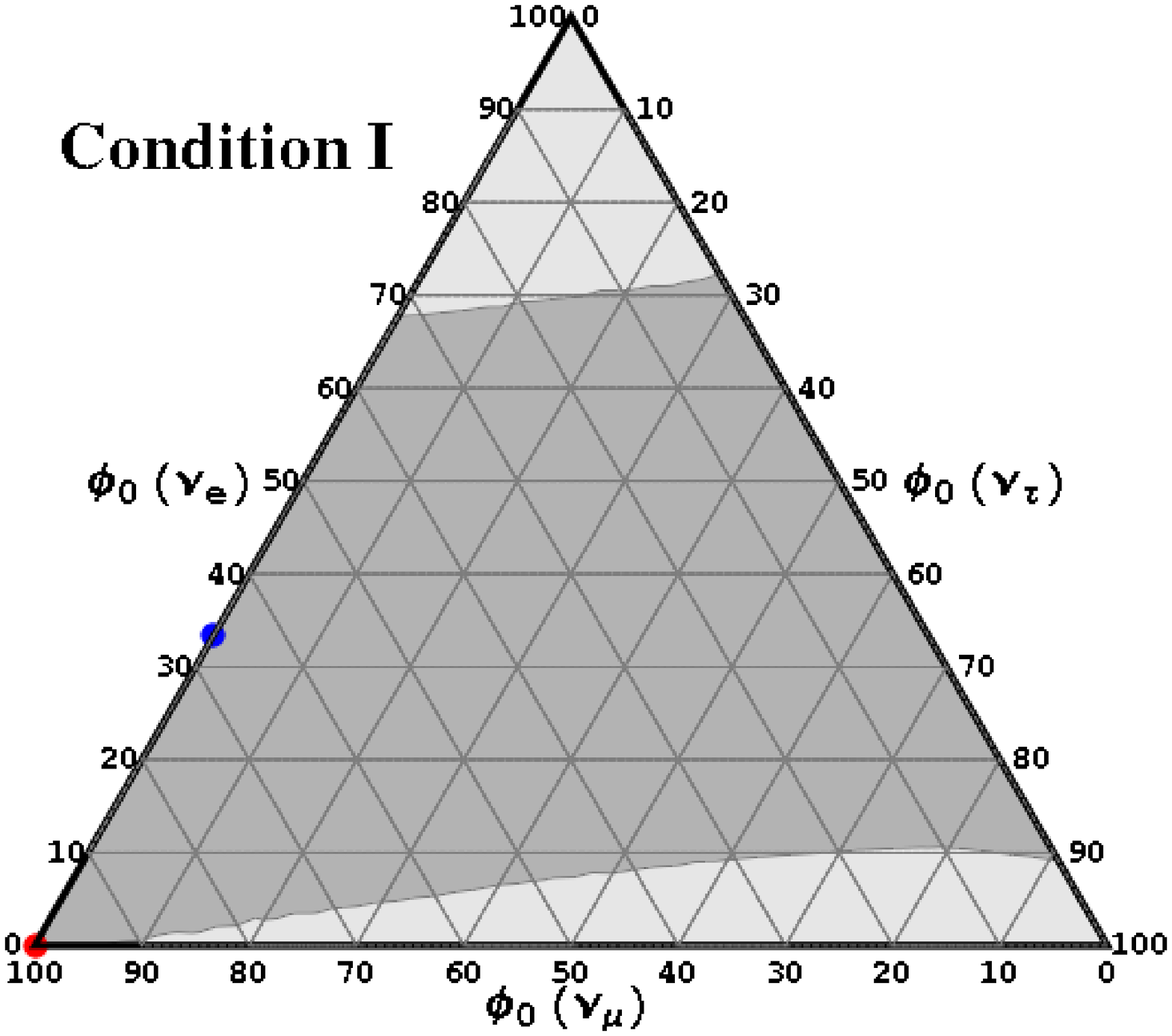}\includegraphics[scale=0.35]{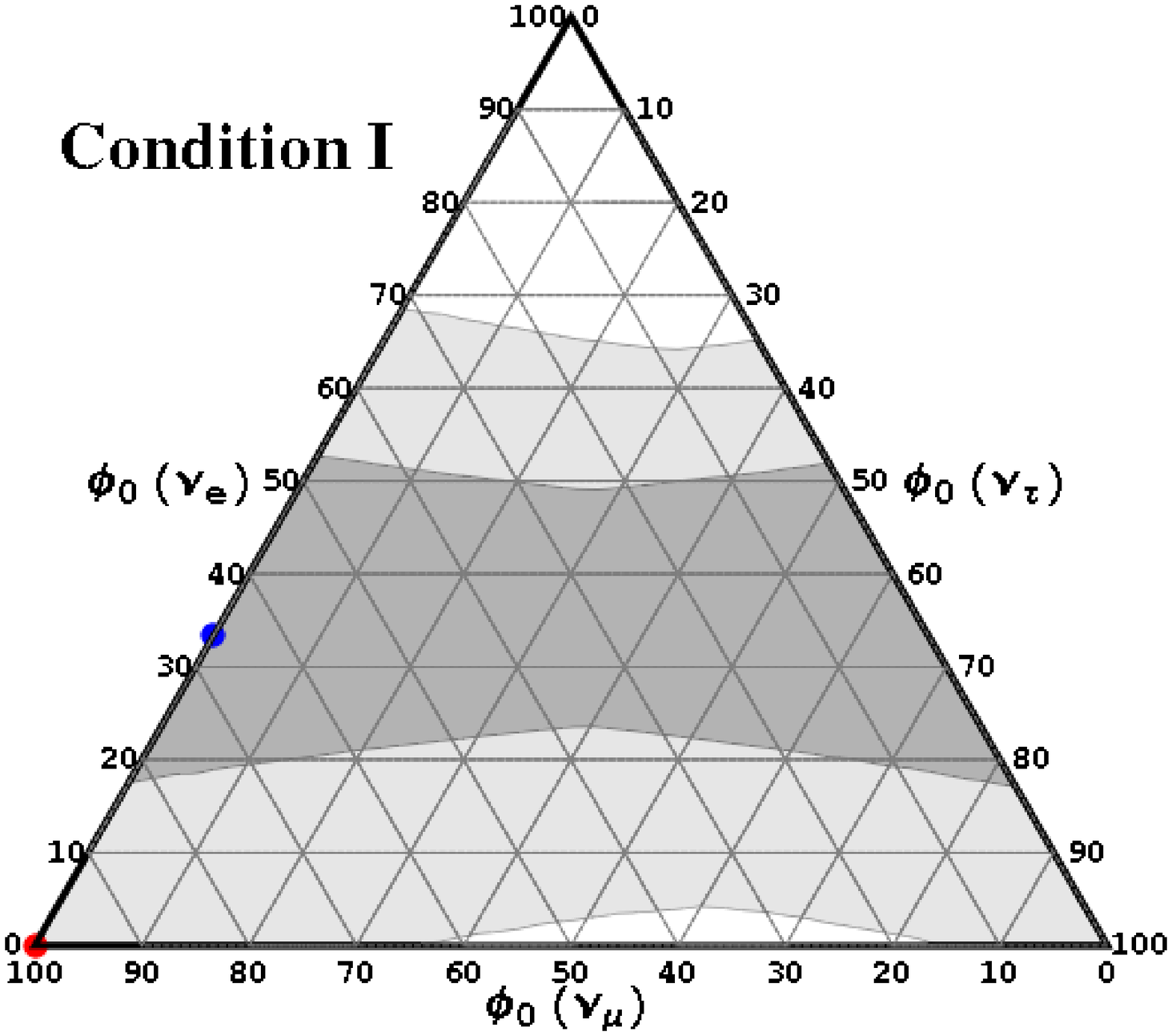}

\caption{ Reconstructed flavor ratios for an input pion source in
the energy regime $E_{\nu}< 33$ PeV. The dark and light shaded areas
denote the 1$\sigma$ and 3$\sigma$ ranges respectively. The left
panel is the result obtained by measuring only $R^{\rm I}$ to the accuracy of
$\Delta R^{{\rm I}}/R^{{\rm I}}=15\%$. The right panel is the result
obtained by measuring both $R^{\rm I}$ and $S^{\rm I}$ with $\Delta
R^{{\rm I}}/R^{{\rm I}}=15\%$ and $\Delta S^{{\rm I}}/S^{{\rm I}}$
related to the former by Poisson statistics. }

\label{fig:low_energy}
\end{figure}

In this energy range, the appropriate flavor discrimination
variables are $R^{\rm I}$ and $S^{\rm I}$ respectively as summarized
in Table II. It has been shown in Ref.~\cite{Lai} that measuring
$S^{\rm I}$ significantly improves the constraint on the initial
neutrino flavor ratio obtained by measuring $R^{\rm I}$ alone.
Taking pion source as the input true source, the effect of measuring
$S^{\rm I}$ is presented in Fig.~\ref{fig:low_energy}.
The left panel of the figure shows the reconstruction result obtained by measuring
only $R^{\rm I}$ to the accuracy of $\Delta R^{{\rm I}}/R^{{\rm
I}}=15\%$ \cite{accuracy}. At $3\sigma$ level, any initial neutrino flavor ratio is
allowed. The right panel is the result obtained by measuring both $R^{\rm I}$ and
$S^{\rm I}$ with $\Delta R^{{\rm I}}/R^{{\rm
I}}=15\%$ and $\Delta S^{{\rm I}}/S^{{\rm I}}$ related to the former
by Poisson statistics. Clearly the further measurement of $S^{\rm
I}$ drastically improves the constraint on initial neutrino flavor
ratio.
\section{Discussions and Conclusions}
To reconstruct the source flavor ratio of ultrahigh energy
neutrinos, it is crucial to have sufficient event numbers and an
effective way of separating $\nu_e$ from $\nu_{\mu}$ and
$\nu_{\tau}$. For $E_{\nu}>33$ PeV, one expects that the cosmogenic
neutrino flux \cite{cosmogenic_early,cosmogenic_update} arising from
GZK interactions \cite{Greisen,ZK} should dominate over those fluxes
originated from astrophysical sources, such as GRB
\cite{Waxman:1998yy,Baerwald:2010fk}. It has been shown that the
newly proposed Askaryan Radio Array \cite{Allison:2011wk} can detect
roughly $50$ cosmogenic neutrino events in 3 years for baseline flux
models such as those discussed in Ref.~\cite{cosmogenic_update}.
This event number implies a nearly $30\%$ accuracy (statistically)
in 3 years or a 15$\%$ accuracy in a decade of data-taking for
determining the flux ratio $R^{\rm II}$. Certainly the efficiency of
flavor discrimination also affects the accuracy for determining
$R^{\rm II}$. Further studies are needed on this aspect.
\begin{figure}
\includegraphics[scale=0.50]{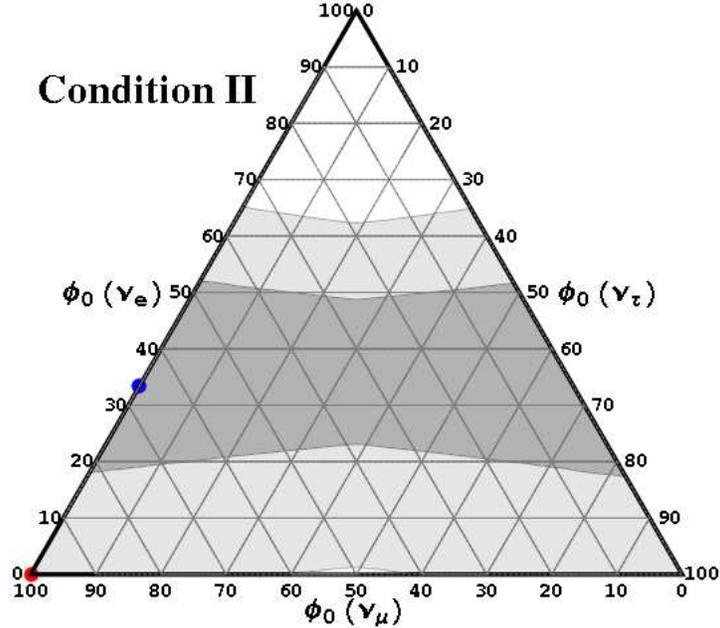}

\caption{Reconstructed flavor ratios for an input pion source with
$\Delta R^{{\rm II}}/R^{{\rm II}}=15\%$ and the uncertainties of
neutrino mixing angles reduced to $50\%$ of those listed in
Eq.~(\ref{angles}). The dark and light shaded areas denote 1$\sigma$
and 3$\sigma$ ranges respectively. The result shown in this figure
is comparable to that in Fig.~\ref{fig:PionOnlyR}}
\label{fig:pion_reduced}
\end{figure}

It is of interest to investigate whether or not a better knowledge
of neutrino mixing parameters will improve the constraint on source
flavor ratio. Taking the pion source as the input true source and
considering the uncertainties of neutrino mixing parameters being
half of those listed in Eq.~(\ref{angles}), we obtain the
reconstructed flavor ratios as shown in Fig.~\ref{fig:pion_reduced}.
One cannot see noticeable difference between this
result and that of Fig.~\ref{fig:PionOnlyR}. It is clear that the
constraint on source flavor ratio solely depends on the accuracy of
measuring $R^{\rm II}$.

In summary, we have argued that the flux ratio $R^{{\rm
II}}=\phi(\nu_{e})/(\phi(\nu_{\mu})+\phi(\nu_{\tau}))$ is a suitable
variable for neutrino flavor discrimination for $E_{\nu}>33$ PeV in
water (ice) Cherenkov and radio wave detectors. In view of recent
development in radio wave array, we studied the reconstruction of
neutrino flavor ratio at the source with $\Delta R^{{\rm
II}}/R^{{\rm II}}$ taken to be $15\%$ and the uncertainties of
neutrino mixing parameters given by Eq.~(\ref{angles}). We have
demonstrated that the further distinction between $\nu_{\mu}$ and
$\nu_{\tau}$ in such a high-energy range does not improve the
constraint on the source flavor ratio. This is a consequence of
approximate $\nu_{\mu}-\nu_{\tau}$ symmetry.

 This work is supported by National Science Council of
Taiwan under Grants No. NSC 96-2112-M-009-023-MY3 (T.C. Liu and G.L.
Lin) and NSC 97-2112-M-239-002 (M.A. Huang) from the National
Science Council of Taiwan.


\begin{thebibliography}{10}
\bibitem{comment} It is understood that $e^{\pm}$ in type-A event releases all its energy into shower while $\tau^{\pm}$ in type-C event does not. Hence, for a given shower energy threshold, the number of type-A event is in general more than that of type-C event for $\phi(\nu_e)=\phi(\nu_{\tau})$. Such a difference could possibly be removed by lowering the shower energy threshold.
\bibitem{accuracy}
For convenience, we have taken $\Delta R^{{\rm I}}/R^{{\rm
I}}=\Delta R^{{\rm II}}/R^{{\rm
II}}=15\%$. However, it is important to note that the accuracies for measuring $R^{\rm I}$ and $R^{\rm II}$ depend on the magnitude of neutrino flux and the neutrino detection technique at each corresponding energy range.
\end{thebibliography}
\end{document}